\documentclass[structabstract]{aa}
\usepackage{graphicx}
\usepackage{txfonts}
\usepackage[usenames]{color}           
\newcommand{\kev}[1]{{#1}}

\newcommand{\eq}[1]{Equation~(\ref{eq:#1})}

\newcommand{\sect}[1]{Section~\ref{sec:#1}}

\newcommand{\fig}[1]{Figure~\ref{fig:#1}}
\newcommand{\figs}[2]{Figures~\ref{fig:#1} and \ref{fig:#2}}

\newcommand{\eg}{\textit{e.g.}, }
\newcommand{\ie}{\textit{i.e.}, }

\begin{document}
   \title{Can we explain non-typical solar flares?}
   
   \author{
      K. Dalmasse,
      \inst{1}
      R. Chandra,
          \inst{2}
      B. Schmieder,
      \inst{1}
       \and
      G. Aulanier
     \inst{1}
          }

\institute{
LESIA, Observatoire de Paris, LESIA, CNRS, UMPC, Univ. Paris Diderot, 5 place Jules Janseen, 92190 Meudon, France\\
\and
Department of Physics, DSB Campus, Kumaun University, Nainital- 263 002, India
             }
             
   \date{}             


\abstract
    {
We used multi-wavelength high-resolution data from ARIES, THEMIS, and SDO 
instruments, to analyze a non-standard, C3.3 class flare produced within the 
active region NOAA 11589 on 2012 October 16. Magnetic flux emergence and 
cancellation were continuously detected within the active region, the latter leading 
to the formation of two filaments.
    }
    {
Our aim is to identify the origins of the flare taking into account the complex 
dynamics of its close surroundings.
    }
    {
We analyzed the magnetic topology of the active region using a linear force-free 
field extrapolation to derive its 3D magnetic configuration and the location of 
quasi-separatrix layers (QSLs) which are preferential sites for flaring activity. 
\kev{Because the active region's magnetic field was nonlinear force-free, 
we completed a parametric study using different linear force-free field 
extrapolations to demonstrate the robustness of the derived QSLs.}
    }
    {
The topological analysis shows that the active region presented a complex 
magnetic configuration comprising several QSLs. The considered data set 
suggests that an emerging flux episode played a key role for triggering the flare. 
The emerging flux \kev{likely} activated the complex system of QSLs leading 
to multiple coronal magnetic reconnections within the QSLs. This scenario accounts 
for the observed signatures: the {two extended} flare-ribbons developed at locations 
matched by the photospheric {footprints} of the QSLs, and were accompanied 
with flare loops that formed {\it above} the two filaments which played no 
important role in the flare dynamics.
    }
    {
This is a typical example of a complex flare that {can a-priori show standard 
flare signatures that} are nevertheless impossible to interpret with any standard 
model of eruptive or confined flare. We find that a topological analysis however 
permitted to unveil the development of such complex sets of flare signatures.
    }

\titlerunning{A non-standard flare on 2012 October 16}
\authorrunning {K. Dalmasse et al.}
   \maketitle
%

\section{Introduction} \label{sec:S-Introduction}

%
%
Solar flares are the most energetic events on the sun. They emit radiation over the 
whole electromagnetic spectrum from $\gamma$-rays to radio wavelengths 
\citep{Shibata99,Shibata11}. Magnetic reconnection is the main process that releases 
energy during the solar flares. This energy is extracted from the magnetic energy that 
is stored in current-carrying fields in the corona. During a flare, energetic particles and 
thermal energy are produced around the reconnection site. They flow down towards 
the lower and denser layers of the solar atmosphere. As a result, coronal emission is 
produced within and around (post) flare loops, and surface brightenings occur along 
so-called flare ribbons, as observed in the ultraviolet (UV) as well as in 
typically-chromospheric wavelengths such as H$\alpha$. Solar flares are usually 
classified into two categories: eruptive or confined. 

%
%
When a flare is associated with a coronal mass ejection (CMEs), either being 
associated with a detectable filament eruption or not, it is an eruptive flare. Those are 
often referred to as two-ribbon flares and long duration events, because they are 
associated with two parallel flare ribbons, that are located on both sides of the polarity 
inversion line (PIL), and that gradually move apart from one another. So as to explain 
the different observational manifestations of eruptive flares such as filament eruptions 
when they are observed, ribbon separations, flare loops formation, and associated 
phenomena, the standard CSHKP flare model was developed in two dimensions 
\citep{Carmichael64,Sturrock66,Hirayama74,Kopp76,Forbes86}. According to this 
model, a current sheet forms in the corona, right below the erupting filament. Magnetic 
field lines sequentially reconnect at this current sheet, resulting in a growing 
(resp. spreading) system of flare loops (resp. ribbons), located below the erupting 
filament. Some 3D extensions to this model have been recently proposed to explain 
observational properties and physical processes, firstly in the form of cartoons 
\citep{Shibata95, Moore01, Priest02} and more recently based on numerical simulations 
\citep{Aulanier12,Kusano12, Janvier13}. 

%
%
The other flares, that are not associated with a CME, are the confined flares. Those 
are {classically} due to loop-loop interactions in the corona, which are induced 
by horizontal motions or flux emergence through the photosphere 
\citep[\eg][]{Gorbachev89,Demoulin97, Hanaoka97, Mandrini97, Schmieder97, Nishio97, Chandra06}. 
Confined flares are usually associated with multiple ribbons. The classical two-dimensional 
picture for the magnetic configuration and reconnection behavior in such flares is that 
of a coronal X-point, at which a current sheet is gradually formed as a result of the 
photospheric motions \citep{Giovanelli47,Heyvaerts77,Syrovatskii81,Low88,Aly97}. 
Magnetic topology  analyses of active regions have played a crucial role in 
understanding the magnetic reconnection processes in 3D in confined flares 
\citep[see review by][]{Demoulin07}. In 2D configuration, the reconnection can occur 
at null points, where the magnetic field vanishes. In 3D, the reconnection can also 
occur at a null point \citep{Masson09}, but also along a separator 
\citep[\eg][]{Longcope05,Parnell10a} or a quasi-separatrix layer 
\citep[QSL, see \eg][]{Demoulin97,Titov02,Aulanier05,Pariat12}. 

%
%
Some atypical flares share several elements common to both the {classical definition of} 
eruptive and confined categories, in particular the existence of two parallel ribbons and several 
other remote ribbons. To the authors' knowledge, three different origins are known 
for these complex events {which, depending on each case, either belong to the eruptive 
or confined flares category.} 
Firstly, they can be due to a failed filament eruption. The 
confinement of the filament by coronal arcades eventually makes it stall in the low 
corona, and eventually reconnect with its restraining arcades 
\citep[\eg][]{Torok05,Guo10,Chen13}. Secondly, they can develop when long-distance 
loop-loop interactions and reconnections are driven by a successful eruption that 
pushes these loops against their neighbors \citep[\eg][]{Maia03,Chandra09}. Thirdly, 
they can appear when two filaments of opposite helicities reconnect with one another 
without merging \citep{Deng02,Schmieder04,DeVore05,Torok11,Chandra11}. 

%
%
Because of their complexity, many atypical flares have {not} been analyzed in great 
details. One could wonder if the usual tools and models that have been developed 
throughout the years are really relevant for all of these complex events. The question 
is more preoccupying than it sounds a priori, since these complex under-looked flares 
may be the most numerous, among all the flares that the Sun produces. 
{We note that the recent paper by \cite{Liu14} was the first topological study that 
started addressing this question. Combining a careful EUV analysis with the QSL method, 
the authors were able to identify their event as being a confined flare associated with 
a failed flux rope eruption.} 
The aim of our paper is {to present and analyze a different but} complex event that 
involved filaments, therefore using the standard flare model and the 
QSL method. Our single event was merely selected because it was observed with two 
independent ground based telescopes, namely THEMIS in Tenerife and ARIES in India. 
It was a C3.3 class flare, that occurred on Oct 16, 2012 in the active region NOAA 11589. 
This region comprised two filaments, that gradually formed and converged, but did not 
merge. 

%
%
The QSL method was first proposed in \cite{Demoulin97}. It is based on the calculation 
of the photospheric footprints of QSLs, from extrapolated magnetic fields. QSLs are defined 
as the narrow volumes within which the magnetic field connectivity has very sharp gradients 
\citep{Priest95}. They are the 3D generalization of separatrices in 2.5D X-points with an 
additional guide field \citep[those were called flipping layers by][]{Priest92}. QSLs are 
preferential sites for the build-up of electric currents and the development of magnetic 
reconnection in general 3D systems. Among many developments, QSLs have been shown 
to play an essential role not only in confined flares, but also in eruptive flares 
\citep{Demoulin96,Savcheva12,Janvier13}, possibly in SEP transport towards Earth 
\citep{Masson12} as well as in twisted flux tubes interacting in solar observations 
\citep{Chandra11}, in numerical simulations \citep{Milano99,WilmotSmith10,Torok11} 
and in laboratory experiments \citep{Lawrence09,Gekelman12}. More details can be 
found in the reviews by \citet{Demoulin06} and \citet{Aulanier11}. So as to conduct the 
QSL method (\ie to plot the photospheric footprints of QSLs), either the norm $N$ of the 
QSL \citep{Demoulin97} or its squashing degree $Q$ \citep{Titov02}, have to be calculated 
at the boundary of the extrapolated fields. Since both $N$ and $Q$ provide a different 
measure for the gradients of the field line connectivity across QSLs, their footprints 
naturally arise as narrow and elongated layers where $N$ or $Q \gg 1$. In this paper, 
we apply the QSL method to NOAA 11589, {by computing the squashing degree, Q, 
at the photospheric level.}

%
%

The paper is organized as follows. \sect{S-Observations} presents the observations, with an 
analysis of the evolution of two filaments in the active region, and the development of the flare. 
The QSL method and the potential role of QSLs in the flare are discussed in \sect{S-Topology}. 
In \sect{S-confined-flare}, we present our interpretation of our results, with an observational 
evidence for the trigger of the flare, and with a conjecture on the sequences of reconnections 
in the calculated QSLs that can account for the complex development of the observed atypical 
flare. Finally, in \sect{S-Summary-Discussion}, we conclude on the important role of the QSL 
method in unveiling the sequence of events that shape complex and atypical flares, even when 
they do not fit the standard model.

%

\section{Observations} \label{sec:S-Observations}

\subsection{Data} \label{sec:S-Data}

Part of the observations of NOAA 11589 presented here was obtained with the 
Atmospheric Imaging Assembly imager \citep[AIA;][]{Lemen12} and the Helioseismic 
and Magnetic Imager \citep[HMI;][]{Schou12} onboard the Solar Dynamic Observatory 
\citep[SDO;][]{Pesnell12} satellite. The AIA instrument observes the Sun over a wide 
range of temperatures from the photosphere to the corona. The pixel size of the AIA 
images is $0.6''$. In this study, we considered the 1600, 304, 193, and 171 \AA\ data. 
The magnetic field in the AR was studied by using the line-of-sight magnetograms 
of the HMI instrument which observes the full disk with a pixel size of $0.5''$.   

We also used ground-based observations of the AR obtained with the indian 
telescope from the Aryabhatta Research Institute of observational Sciences (ARIES), 
and with the french T\'elescope H\'eliographique pour l'Etude du Magn\'etisme et des 
Instabilit\'es Solaires (THEMIS). The 15-cm f/15 Coud\'e telescope of the ARIES, 
operating in Nainital (India), observes in the H$\alpha$ line with a spatial-resolution 
of $0.58''$. The THEMIS telescope, operating in Tenerife (Canary Islands), allows to 
simultaneously map the H$\alpha$ emission and the full Stokes parameters in the 
Fe $6302.5$ \AA\ of a field-of-view of about $240'' \times 100''$ in one hour.

  \begin{figure*}
   \centerline{\includegraphics[width=0.85\textwidth,clip=]{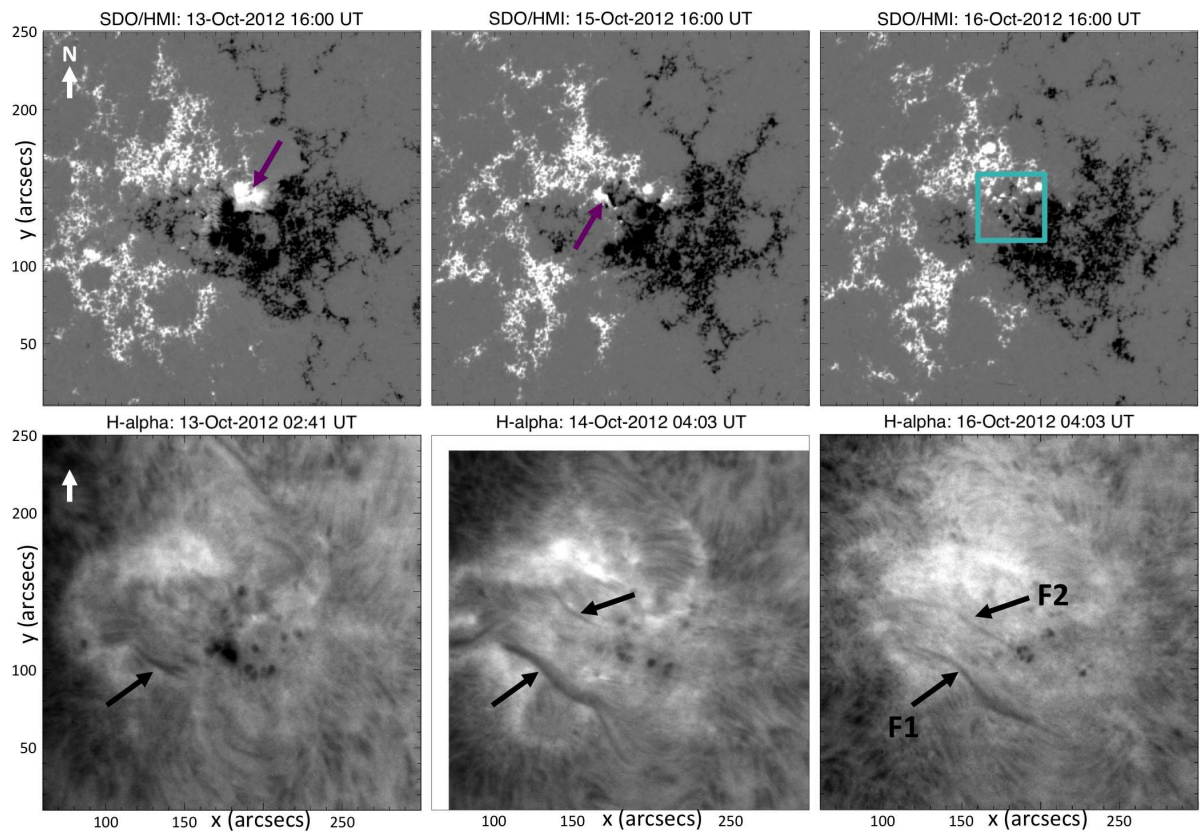}
              }
              \caption{Evolution of active region NOAA 11589 during its disk passage before the eruption on 2012 October 16. Top: Evolution of the longitudinal magnetic field observed by SDO/HMI. White/black are positive/negative polarities. The field strength is saturated at 500 Gauss. The violet arrows indicate significant emerging fluxes on October 13 and 15. {The cyan rectangle highlights the region where recurring magnetic flux emergence occurred on October 16 and likely triggered the studied C3-class flare (see \sect{S-Flare}).} \kev{The temporal evolution of the magnetograms is available as a movie in the online edition.} Bottom: Development of filaments in H$\alpha$ observed by ARIES telescope. The locations of two observed filaments F1 and F2 are indicated by {black} arrows. The {white} arrow indicates the north direction.
                      }
   \label{fig:Fig-Mgm-Halpha} 
   \end{figure*} 

\subsection{Evolution of the photospheric magnetic field} \label{sec:S-Bfield-evolution}

The AR NOAA 11589 appeared at the heliographic coordinates N13 E61 on 2012 
October 10. The AR appeared as two large-scale, decaying magnetic polarities. 
It presented a $\beta$ magnetic configuration which evolved towards a 
$\beta \gamma \delta$ configuration on October 16. During its on-disk passage, 
the AR produced 20 C-class flares.

The evolution of the AR during its on-disk passage presented localized magnetic 
flux emergence episodes together with large-scale magnetic flux cancellation as 
displayed in \fig{Fig-Mgm-Halpha} (top row). The episodic emerging flux events 
occurred within the north of the central part of the AR. The violet arrows in 
\fig{Fig-Mgm-Halpha} highlight two of these emerging flux events which occurred 
on October 13 and 14.

The magnetograms evolution also presents traces of large-scale magnetic flux 
cancellation. In particular, we can see that the positive polarity, pointed by the 
violet arrow in the magnetogram of October 13, was progressively cancelled out. 
On October 16, this positive polarity had almost vanished. The large-scale 
flux cancellation is also particularly well observable in the central part of the AR, 
on the east part of the negative polarity. Indeed, it shows that the easternmost 
part of the negative polarity moved towards the east and progressively cancelled 
out with the positive polarity.

\subsection{Evolution of the two active region filaments} \label{sec:S-Filaments-evolution}

The large-scale magnetic flux cancellation observed in the central part of the AR 
led to the formation of two filaments 
\citep[\eg][]{vanBallegooijen89,Antiochos94,Martens01,Wang07}. 
The evolution of these filaments in H$\alpha$ is presented in \fig{Fig-Mgm-Halpha} 
(bottom row). The formation of the first filament started on October 13 
(see \fig{Fig-Mgm-Halpha}). The filament appeared on the southern part of the AR, 
and progressively evolved towards the thick and elongated filament labeled F1 in 
the H$\alpha$ image of \fig{Fig-Mgm-Halpha}. The second filament appeared on 
October 14 in the center of the AR, and progressively evolved towards the filament 
labeled F2.

\begin{figure}
   \centerline{\includegraphics[width=0.49\textwidth,clip=]{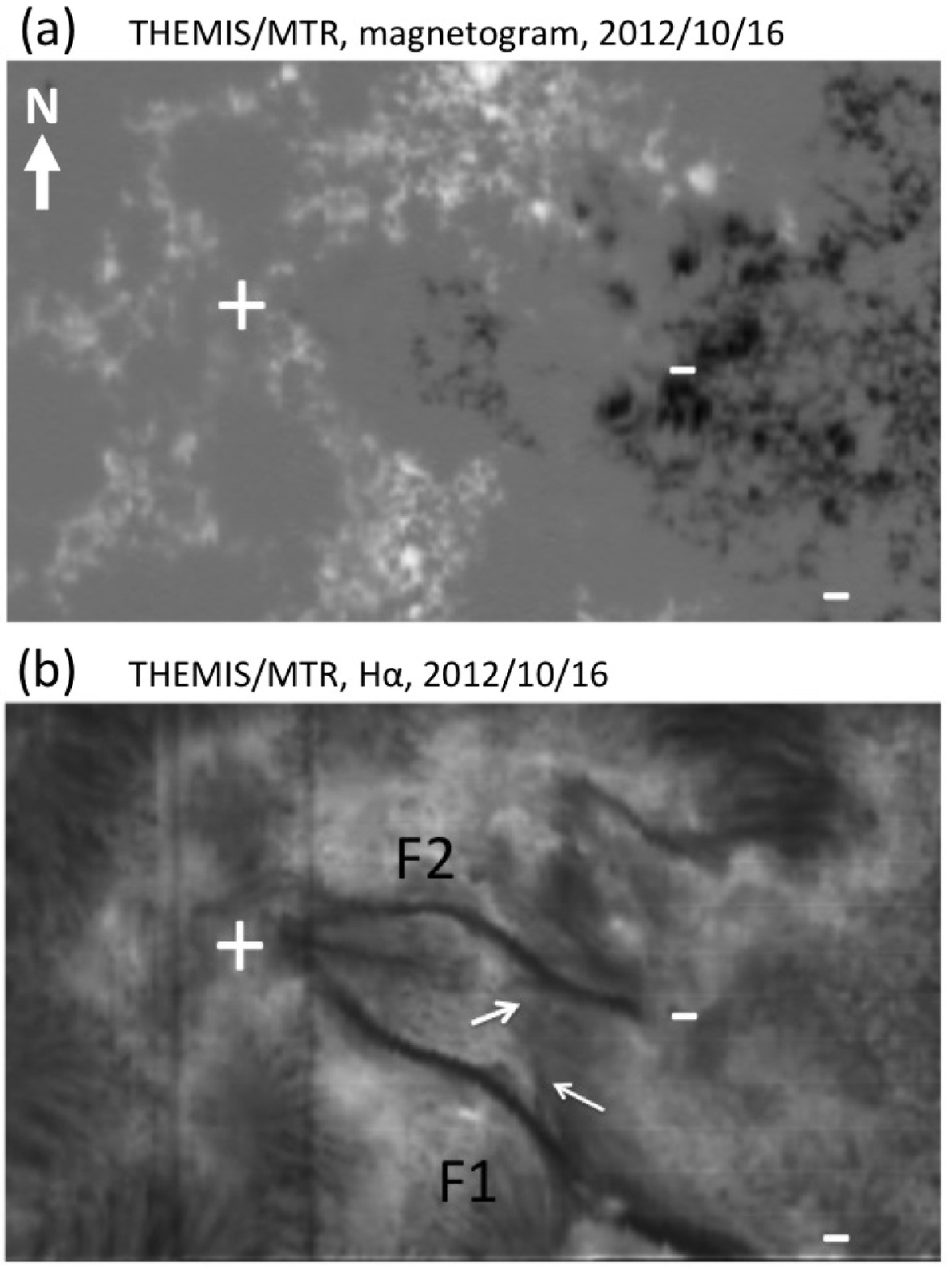}
              }
              \caption{Active region NOAA 11589 observed on 2012 October 16 by THEMIS/MTR between 08:02 and 09:02 UT. Top: Longitudinal magnetic field. Bottom: H$\alpha$ map showing the two recently formed filaments {of \fig{Fig-Mgm-Halpha}.} The white arrows indicate the barbs used to infer the filaments chirality. Filament F1 is a dextral filament, F2 is  a sinistral filament. The $+$ and $-$ signs indicate the magnetic field polarity of each end of the filaments. The field of view covers $\sim 175'' \times 100''$. The {white} arrow indicates the north direction.
                      }
   \label{fig:Fig-THEMIS-data} 
   \end{figure}

Using {the H$\alpha$ data from THEMIS} (\fig{Fig-THEMIS-data}), we 
{were able to derive} the chirality of the filaments based 
on \cite{Aulanier98} and \cite{Mackay10}. In \fig{Fig-THEMIS-data}, one of 
the barbs of filament F1, highlighted by the southern white arrow, indicates that 
the filament was dextral. In addition, the filament F1 had its easternmost end 
rooted in the positive polarity and its westernmost end rooted in the negative 
polarity. This indicates that its axial field was pointing towards the south-west. 
Regarding the position of the positive polarity compared with the negative polarity 
in this region {(\fig{Fig-Mgm-Halpha}),} it follows that the filament was dextral 
and thus had a negative helicity, which agrees with the orientation of the filament 
barbs. We note that the filament F1 thus obeyed the hemispheric chirality rule, 
{according to which} most of the filaments 
of the northern hemisphere have a dextral chirality \citep[\eg][]{Pevtsov03}. Based 
on the same analysis, we found that the chirality of filament F2 was sinistral. The 
filament F2 thus had a positive helicity. Hence, F2 did not obey the hemispheric 
chirality rule. We thus conclude that NOAA 11589 possessed a mixed magnetic 
helicity, with positive magnetic helicity in its northern part, and negative  magnetic 
helicity in its southern part (see also \sect{S-Bextrapol}).

The evolution of these two filaments shows that the northern footpoints of both 
filaments converged towards each other without merging. This is in agreement 
with previous numerical simulation \citep[\eg][]{DeVore05,Aulanier06} and 
observational studies \citep[\eg][]{Martin98,Schmieder04,Chandra10,Torok11,Chandra11} 
showing that the merging of two filaments strongly depends on their chirality and 
their relative orientation. In particular, the presented filaments evolution would be 
equivalent to {\it Experiment 2} of \citeauthor{DeVore05} (\citeyear{DeVore05}, see their Fig. 8). 
Thus, the filaments did not have the opportunity of merging probably because 
their axial field was oriented in opposite direction along the PIL.

\subsection{The 2012 October 16 flare} \label{sec:S-Flare}

On 2012 October 16, the AR was located at heliographic coordinates N13 W11. 
On that day, the AR produced a C3.3/1F class flare. According to the GOES 
instruments, the flare started around 16:12 UT, peaked at 16:27 UT, and ended 
around 16:39 UT. The flare signatures were visible in the different wavelengths 
observed by the SDO. The 94 \AA\ data from the SDO/AIA indicates that the flare 
was initiated in the northern part of the AR where magnetic flux emergence was 
often detected (see the violet arrows in \fig{Fig-Mgm-Halpha}). 

\begin{figure*}
   \centerline{\includegraphics[width=0.99\textwidth,clip=]{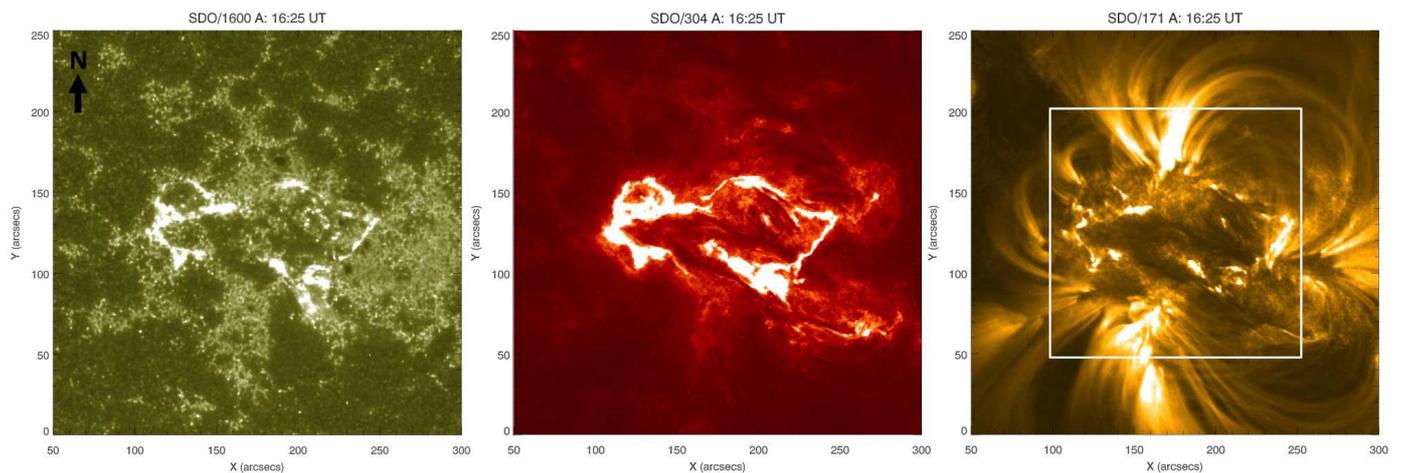}
              }
              \caption{Flare signatures observed by SDO/AIA on 2012 October 16, at 1600 \AA\ (left), at 304 \AA\ (middle), and at 171 \AA\ (right). The black arrow indicates the north direction. The white square indicates the field-of-view of \fig{Fig-193-pfl}. \kev{The temporal evolution of AIA 1600 \AA, 304 \AA, and 171 \AA\ images is available as a movie in the online edition.}
                      }
   \label{fig:Fig-flare} 
   \end{figure*}

\fig{Fig-flare} displays the flare signatures at 1600 and 304 \AA\ during the 
maximum phase of the flare. These signatures present a similar morphology 
in both wavelengths. During the flare evolution, the data show the beginning 
of small, localized brightenings appearing on the north, east, and south parts 
of the AR. The eastern brightening, which was also the most distinguishable, 
progressively enhanced and expanded towards the west direction. It formed 
within the positive polarity, and eventually developed into the eastern ribbon 
of \fig{Fig-flare}. The northern brightening, which was the less distinguishable, 
expanded in both the east and west directions. It developed into the northern 
ribbon {of \fig{Fig-flare}} which formed within the positive polarity. {This 
northern ribbon expanded and eventually merged with the eastern ribbon, forming 
a single, extended ribbon within the positive polarity of the AR.} 
The southern brightening, {which formed within the negative 
polarity,} expanded towards the north-west direction, forming the {extended} 
southern ribbon. {Overall, the observations show that the flare-ribbons 
developed into two single, extended ribbons that formed around both filaments, 
one ribbon within the positive polarity, the other within the negative polarity. We note 
that such ribbons are compatible with the two typical flare-ribbons associated with 
the classical eruptive and confined flares involving the presence of a filament. 
Finally,} the observations indicate that, at the {extended} southern ribbon, 
another brightening developed towards the south-west direction between 16:14 
and 16:39 UT. This brightening was probably related to plasma ejection. 
 
\begin{figure*}
   \centerline{\includegraphics[width=0.88\textwidth,clip=]{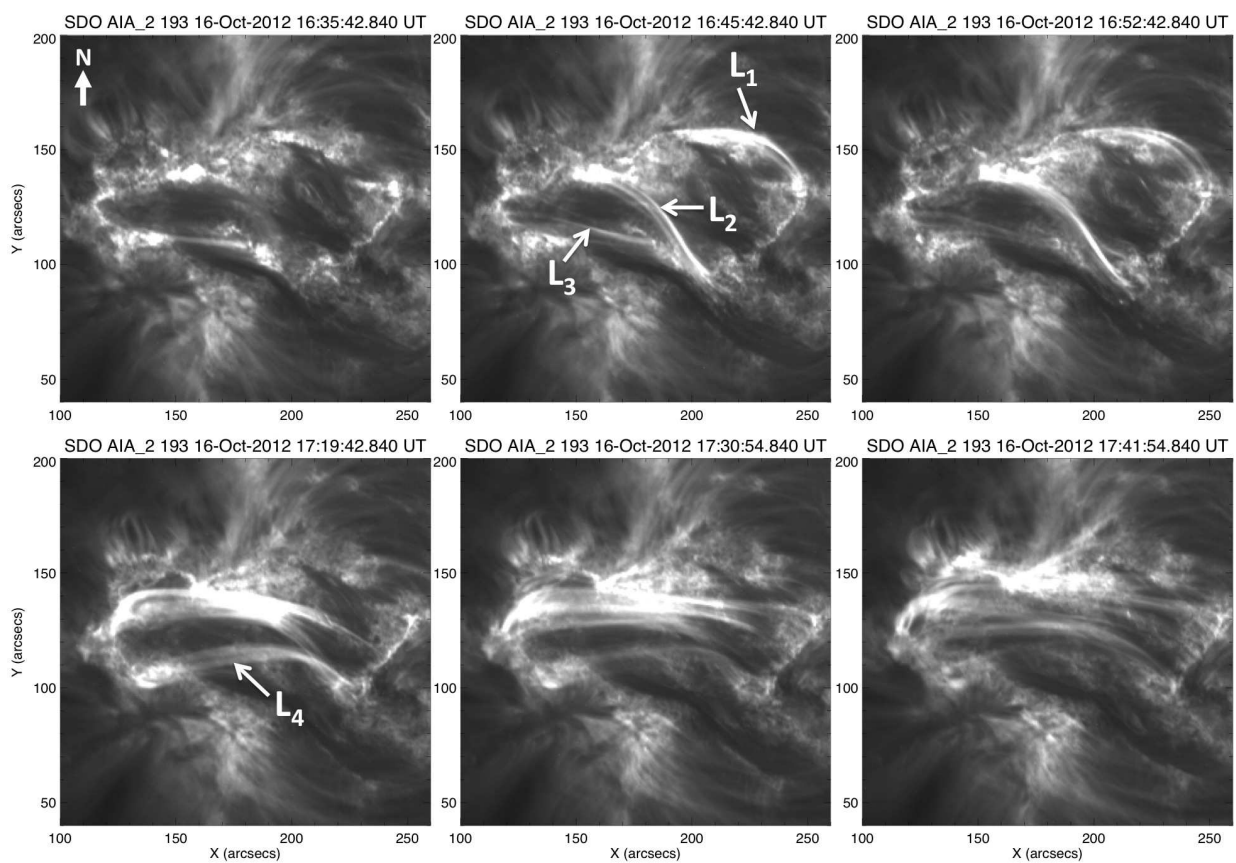}
              }
              \caption{Flare signatures observed by SDO/AIA on 2012 October 16 at 193 \AA. On the top-left panel, the white arrow indicates the north direction.
                      }
   \label{fig:Fig-193-pfl} 
   \end{figure*}

\fig{Fig-193-pfl} presents the evolution of the flare signatures during the decay 
phase at 193 \AA. In this figure, we clearly see the formation of post-flare loops 
joining the {two extended} flare-ribbons displayed in \fig{Fig-flare}. From the 
AR evolution at 193 and 171 \AA, we found that the first post-flare loops developed 
in the northern part of the AR. One of these northern post-flare loops is labeled 
$L_1$ in \fig{Fig-193-pfl}. This post-flare loop was quickly followed by the formation 
of post-flare loops $L_2$ and $L_3$ within the central part of the AR. These post-flare 
loops were then followed by the formation of $L_4$, and a bulk 
of post-flare loops in the central part of the AR.

According to the CSHKP model, both eruptive and confined flares --- involving the 
presence of a filament --- should be associated with the formation of hot post-flare 
loops {\it below} the erupting filament, whether its eruption succeeds or fails 
 \citep[see also][]{Schmieder95,Schmieder96,Shibata11,Aulanier12}. 
Interestingly, we find that the post-flare loops formed {\it above} the filaments. 
Furthermore, the observations indicates that none of the two filaments seemed 
to be neither disturbed nor erupting during or after the flare. These {two} features 
are not consistent with any standard model of eruptive or confined flare. 
{It follows that the two extended flare-ribbons associated with the flare can neither 
be explained by a successful, nor a failed, filament eruption.} A topological 
analysis is then required to build-up a plausible flare scenario that explains the 
observed flare dynamics and its associated signatures.

%

\section{Magnetic topology of the active region} \label{sec:S-Topology}

\subsection{Magnetic field extrapolation} \label{sec:S-Bextrapol}

The topological analysis of AR 11589 magnetic field requires the knowledge of the 
magnetic field in the coronal volume containing the AR. In practice, the coronal 
magnetic field can be estimated from linear 
\citep[\eg][]{Nakagawa72,Alissandrakis81,Demoulin89} or 
nonlinear \citep[see reviews by][and references therein]{Wiegelmann12,Regnier13} 
force-free field extrapolations (LFFF or NLFFF), defined by
    \begin{equation} \label{eq:Eq-LFFF-NLFFF}
      \nabla \times  \vec{B}= \alpha \vec{B} \,,
    \end{equation}
using photospheric data as a bottom boundary condition. In \eq{Eq-LFFF-NLFFF}, 
the force-free parameter, $\alpha$, is uniform in space for LFFF extrapolations, and 
is constant along each elemental flux tubes for NLFFF extrapolations.

Recent studies have shown that NLFFF extrapolations are becoming more and more 
reliable for inferring the coronal magnetic field from photospheric vector magnetograms 
\citep[\eg][]{Schrijver08,Canou10,Valori12,Wiegelmann12b,Guo12,Jiang13b}. 
Because the EUV data show that AR 11589 was formed of filaments of opposite chirality 
(see \fig{Fig-THEMIS-data}) and loops of opposite $\alpha$-values 
(see \eq{Eq-LFFF-NLFFF}), one may want to consider NLFFF extrapolations to study 
the topology of the AR.

However, there are two reasons for not considering such extrapolation models in 
the present study. First, the filaments were located in the plage regions, hence, 
where the magnetic field is weak and the photospheric electric currents, and local 
$\alpha$-values, are not well measured. This would {tend to give a nearly 
potential magnetic field within these regions, which would} prevent from retrieving 
the filaments {in an NLFFF extrapolation} 
\citep[\eg][]{McClymont97,Leka99,Wiegelmann04}. The second reason is given 
by the EUV data showing that none of the filaments seemed to be affected by 
the evolution of the flare. Indeed, both filaments were still present with the same 
shape before and after the flare. In addition, the EUV data show that the post-flare 
loops were formed {\it above} the filaments contrary to what is expected from the 
CSHKP model (see \sect{S-Flare}). Together, 
these observations {\it a priori} suggest that the flare mechanism only involved the 
magnetic field surrounding the filaments, and not the magnetic field of the filaments. 
It is therefore possible to focus the topological analysis of AR 11589 on its global 
magnetic field using simple LFFF extrapolations.

We thus used \eq{Eq-LFFF-NLFFF} with a spatially uniform $\alpha$ to perform 
a set of LFFF extrapolations. The extrapolations were achieved using the method 
described in \cite{Alissandrakis81} for {$5$ distinct LFFF, such that 
$\alpha=[-7,-3.5,0,3.5,7] \times 10^{-3} \ \textrm{Mm}^{-1}$.} 
The method uses fast Fourier transform (FFT) to solve the Helmoltz's equation for 
a LFF magnetic field of force-free parameter $\alpha$. 
{The four side boundary conditions are therefore periodic. There is 
no top boundary condition because the unphysical eigenmodes that increase with 
height are discarded.} 
The magnetogram used as 
the bottom boundary condition ($z=0$) for the extrapolation covers a domain of 
$368 \times 255 \ \textrm{Mm}^2$ {and was taken at 15:00 UT, \eg about one 
hour before the beginning of the flare. Due to the fact that the magnetic field evolves 
only weakly during several days, the exact choice of the magnetogram is not 
determining.}

\begin{figure}   
   \centerline{\includegraphics[width=0.49\textwidth,clip=]{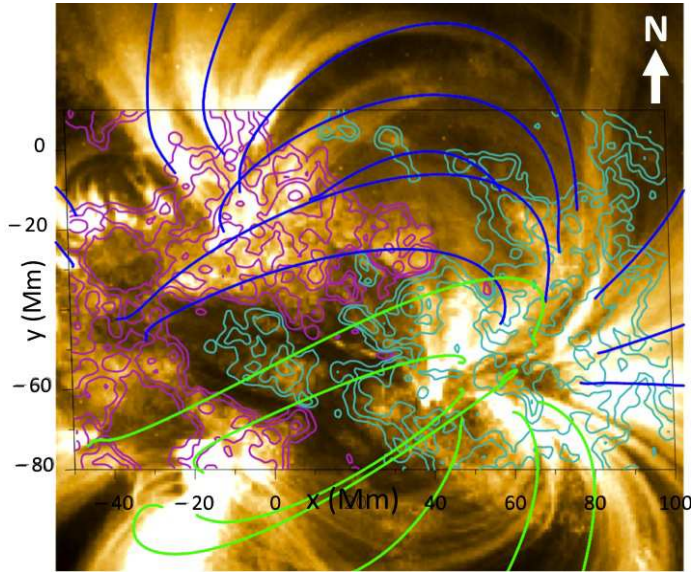}
              }
              \caption{{Zoom on} NOAA 11589 at 15:00 UT on 2012 October 16, {observed with SDO/AIA at 171 \AA\, and overplotted with selected magnetic field lines from the extrapolation ($\alpha = 7 \times 10^{-3} \ \textrm{Mm}^{-1}$). Blue/green lines are magnetic field lines which give a good/poor match with the AR's coronal loops. Solid purple/cyan lines display isocontours of the photospheric magnetic field, $B_z = [30,100,300,1000]$ Gauss. The white} arrow indicates the north direction.
                      }
   \label{fig:Fig-Bz-AIA171}
   \end{figure}

The extrapolations were performed using a 
$xy$-domain roughly twice larger in each direction --- padded with zeros --- in order 
to limit aliasing effects. We {extrapolated the magnetic field up} to $z=2000 \ \textrm{Mm}$, 
leading to an extrapolation domain covering $700^2 \times 2000 \ \textrm{Mm}^3$ 
on a non-uniform grid containing $1024^2 \times 351$ points. Within the set of 
performed extrapolations, we kept the extrapolation giving the best match with 
the northern loops of the AR because this is the region where the flare was initiated 
according to the SDO/AIA 94 \AA\ data. {Using the metrics introduced in \cite{Green02}, 
we found that} the force-free parameter for this extrapolation 
is $\alpha = 7 \times 10^{-3} \ \textrm{Mm}^{-1}$. 
\fig{Fig-Bz-AIA171} displays selected field lines of the magnetic field of this extrapolation 
in the central part of the AR, {plotted over the SDO/AIA 171 \AA\ data.}

\subsection{QSLs in the active region} \label{sec:S-AR11589-QSLs}

\subsubsection{QSLs and flare-ribbons} \label{sec:S-QSLs-shape}

\begin{figure*}
   \centerline{\includegraphics[width=0.99\textwidth,clip=]{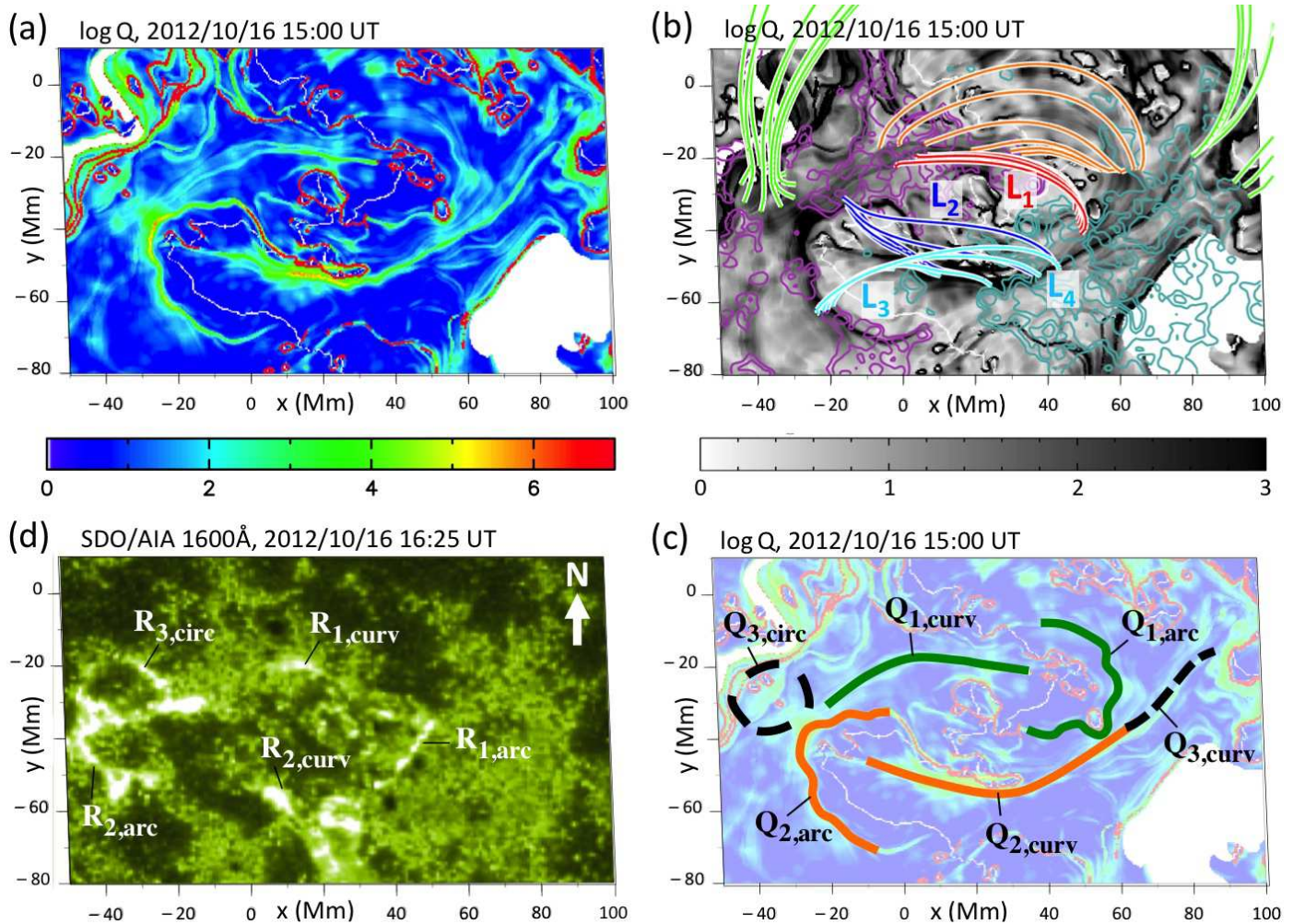}
              }
              \caption{{Zoom on} NOAA 11589. (a) Photospheric mapping of QSLs {from the computation of the squashing degree, $Q$. White regions are related to magnetic field lines which are open at the scale of the {extrapolation domain,} and where $Q$ is not computed.} (b) Selected magnetic field lines and (c) photospheric footprints of the identified QSLs {plotted over the photospheric $Q$-map. The field-lines labeled $L_{i=\{1,2,3,4\}}$ indicate possible candidates for the four post-flare loops labeled in \fig{Fig-193-pfl}.} (d) Flare-ribbons {labelled with respect to the identified QSLs footprints.} The white arrow indicates the north direction.
                      }
   \label{fig:Fig-QSLs-AIA1600}
   \end{figure*}

The computation of {the squashing degree,} $Q$, in the extrapolation domain was 
performed using method 3 of \cite{Pariat12}. \fig{Fig-QSLs-AIA1600}a displays the 
photospheric mapping of QSLs by showing $\log Q$ at $z=0$. Plotting magnetic field 
lines over the $\log Q$ map, we identified three QSLs connected to each other 
{(see \fig{Fig-QSLs-AIA1600}).} The value of $Q$ in these QSLs is typically about 
$10^{3}-10^{4}$ which is indicative of strong connectivity gradients. {For clarity, these 
three QSLs are highlighted and labeled $Q_{i}$ ($i=\{1,2,3\}$) in \fig{Fig-QSLs-AIA1600}b. 
They are respectively compared with the three identified ribbon-systems, $R_{i}$, 
in \fig{Fig-QSLs-AIA1600}c.}

At this point, it must be re-emphasized that QSLs depend on the magnetic field 
connectivity \citep[\eg][]{Demoulin96}, which depends on the extrapolation 
assumptions. This means that extrapolations with different assumptions may lead to 
different QSLs. In some cases, these QSLs could even disappear. For consistency, 
we thus reconsidered all the other extrapolations performed, {\ie 
$\alpha = [-7,-3.5,0,3.5] \times 10^{-3} \ \textrm{Mm}^{-1}$,} and we computed 
{the squashing degree} for all of them {(see \fig{Fig-Qmaps-Online}).}

  \onlfig{
  \begin{figure*}
   \centerline{\includegraphics[width=0.99\textwidth,clip=]{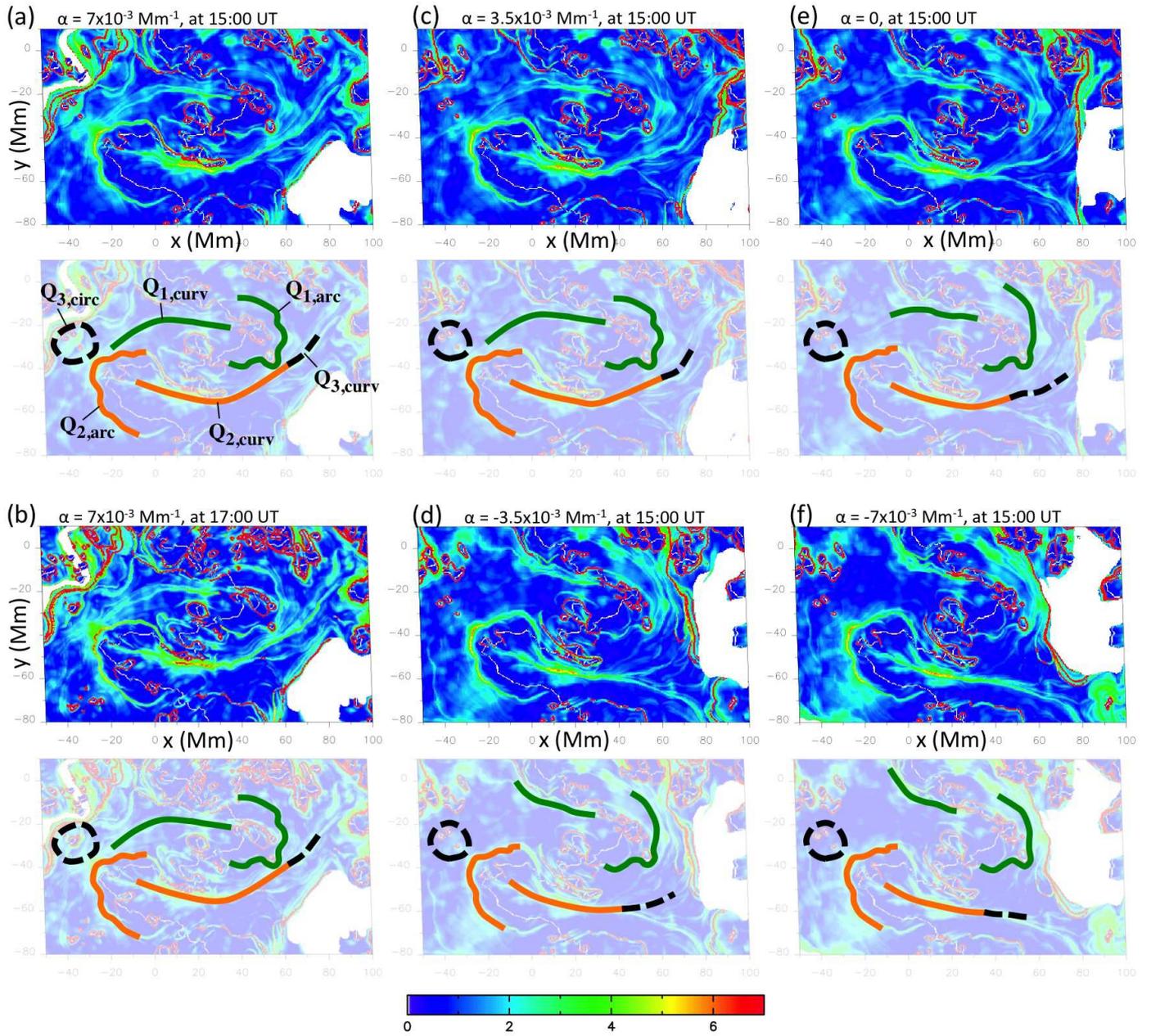}
              }
              \caption{Photospheric mapping of the QSLs of NOAA 11589, from the computation of the squashing degree, $Q$, for all our LFFF extrapolations. (a, c, e, d, f) $\sim 1$ hour before the flare, at 15:00 UT for $\alpha = [7, 3.5, 0, -3.5, -7] \times 10^{-3} \ \textrm{Mm}^{-1}$. (b) $\sim 30$ minutes after the flare, at 17:00 UT, for $\alpha = 7 \times 10^{-3} \ \textrm{Mm}^{-1}$ (\ie the value considered in this paper).
                      }
   \label{fig:Fig-Qmaps-Online}
   \end{figure*} 
   }

  \onlfig{
  \begin{figure*}
   \centerline{\includegraphics[width=0.99\textwidth,clip=]{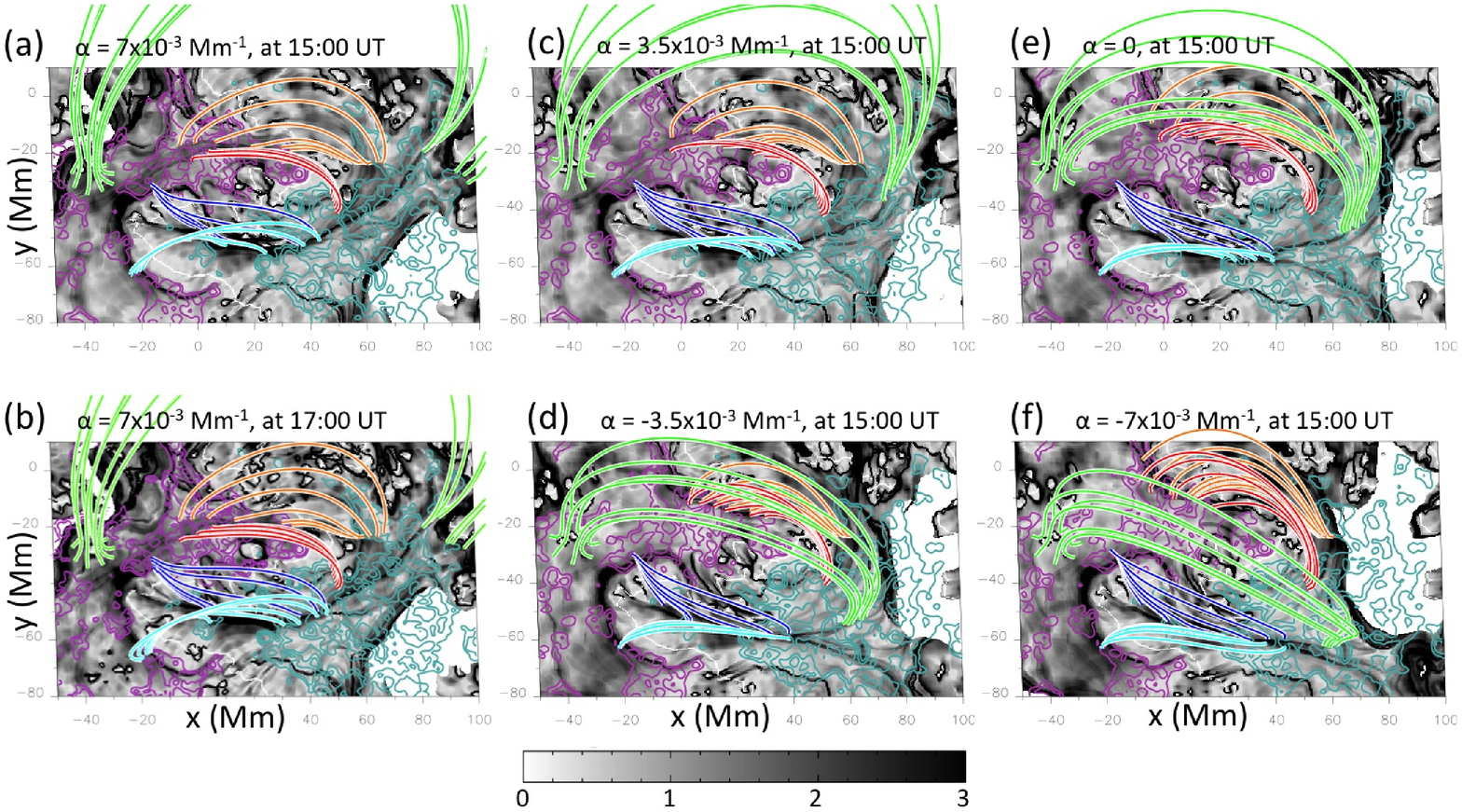}
              }
              \caption{Selected field lines belonging to the three QSLs of NOAA 11589 identified in \fig{Fig-QSLs-AIA1600}b, for the same extrapolations as in \fig{Fig-Qmaps-Online}. Red/orange, dark/light-blue, and green field lines respectively belong to $Q_1$, $Q_2$, and $Q_3$. The grey-scale displays the photospheric map of the squashing degree, $Q$. Solid purple/cyan lines are isocontours of the photospheric vertical magnetic field, $B_z = [150,300,600]$ Gauss.
                      }
   \label{fig:Fig-QSLs-FLs-Online}
   \end{figure*} 
   }


The photospheric footprints of QSLs together with magnetic field lines plotting 
revealed that these three QSLs are reliable 
{(see \figs{Fig-Qmaps-Online}{Fig-QSLs-FLs-Online}).} Indeed, they are present 
in each considered LFFF extrapolations with similar shapes {and locations,} 
meaning that they are topologically robust structures. {There are} only few 
differences that lie on the shapes and intersections of the QSLs footprints. 
{In particular, \figs{Fig-Qmaps-Online}{Fig-QSLs-FLs-Online} show that while 
$Q_2$ and $Q_3$ are always connected regardless of the value of the force-free 
parameter, $Q_1$ and $Q_2$ are solely connected when the force-free parameter 
of the LFFF extrapolation is positive or null. 
From the photospheric mapping of $Q$ (see \fig{Fig-Qmaps-Online}), it is 
clear that only LFFF extrapolations with a positive (or null) force-free parameter 
display QSLs footprints which have a morphology that is compatible with the flare-ribbons 
shown \figs{Fig-flare}{Fig-QSLs-AIA1600}. 
These two figures further justify the use 
of a positive force-free parameter to analyze the topology of the AR's magnetic field, 
and our choice to consider the extrapolation giving the best match with the northern 
coronal loops where $Q_1$ and the trigger of the flare were located. 
Among our LFFF extrapolations, we found that} 
the QSLs from the $\alpha = 7 \times 10^{-3} \ \textrm{Mm}^{-1}$ extrapolation give 
the best match with the flare-ribbons shape (see \fig{Fig-QSLs-AIA1600}c). 
{We emphasize that a magnetic field 
extrapolation performed about 30 minutes after the flare, using 
$\alpha = 7 \times 10^{-3} \ \textrm{Mm}^{-1}$, further shows that the three identified 
QSLs were also temporally robust because they subsisted throughout the duration 
of the flare (see panels (a) and (c) of \figs{Fig-Qmaps-Online}{Fig-QSLs-FLs-Online}).}

Together with magnetic field lines plotting, \fig{Fig-QSLs-AIA1600}a allows to 
distinguish between two double {C-shaped QSL footprints,} $Q_{\{1,2\}}$, and a 
circular-like QSL, $Q_{3}$, in agreement with the three flare-ribbons, $R_{i}$ 
{(see also \fig{Fig-QSLs-FLs-Online}a).} 
A few discrepancies are found between the QSLs footprints and the flare-ribbons 
shape and location, {which results in a rather poor overlay (not shown here).} 
We found the main discrepancies in the identification of 
$Q_{3,curv}$, and in the relative positions of $Q_{2}$ and $R_{2}$. The first is 
related to the difficulty of distinguishing $R_{3,curv}$ from $R_{1,arc}$ and 
$R_{2,curv}$ in the AIA 1600 \AA\, images while it is possible in the extrapolation. 
The observations tend to suggest that, in the real configuration, $Q_{1,arc}$, 
$Q_{2,curv}$ and $Q_{3,curv}$ are more entangled than in the extrapolation. 
The second is related to the deformation of $R_{2,arc}$ compared with $Q_{2,arc}$, 
and the displacement of $R_{2,curv}$ compared with $Q_{2,curv}$. The extrapolation 
shows that $Q_{2,arc}$ is much closer to the PIL than suggested by the corresponding 
flare-ribbon. Also, $Q_{2,curv}$ is very close to the PIL of the northern filament while 
the associated ribbon locates it more in the central part between the two filaments.

It is arguable that all these discrepancies are related to the assumption we made 
by only considering the global magnetic field of the AR and extrapolating it in LFFF. 
Indeed, such a hypothesis does not allow to model the highly-stressed filament 
magnetic fields and their close surroundings. This probably results in local 
modifications of the connectivity of magnetic field lines, which are responsible 
for the deformation and displacement of the QSLs in our extrapolation, as compared 
with the shape and location of the flare-ribbons. Nevertheless, distinctive 
discrepancies between QSLs footprints and flare-ribbons can also be found in NLFFF 
extrapolations. Indeed, this clearly appears in the atypical flare studied by \cite{Liu14}, 
as can be seen in their Figures 7(d) and 7(e). We thus conjecture that such 
mismatches between QSLs footprints and flare-ribbons are more generally inherent 
to \kev{the force-free model of choice.}

Despite the aforementioned 
discrepancies, we find a good qualitative agreement between the QSLs footprints 
and the flare ribbons of our studied event. This match validates the use of a simplified 
LFFF model to study the topology of AR 11589 and relate it to the origin of the flare.

Finally, \fig{Fig-QSLs-AIA1600}a further exhibits two types of very-high $Q$-regions: 
the long red stripes {closed to the open-field regions (white areas in the $Q$-map)} 
at the East/West edges of the AR, and the red segments and 
round-shapes. The first are due to the aliasing from the periodic boundary conditions 
and are spurious. The second are due to very low-altitude null-points located above 
small parasitic polarities. These small QSLs may sustain magnetic reconnection, and 
lead to small-scale jets and bright-points. However, they are unrelated to the flare 
because their field lines do not intersect the QSL system $Q_{1,2,3}$. We therefore 
ignore them in our analysis.

\subsubsection{A complex {interlinked} topology} \label{sec:S-Complex-topo}

\fig{Fig-QSLs-AIA1600}b displays a cartoon of the inferred magnetic topology {plotted 
over the photospheric $Q$-map.} 
It comprises the two double {C-shaped} QSLs (green and orange QSLs) that resemble 
the QSL of the quadrupolar magnetic configuration from \cite{Titov02} or \cite{Aulanier05}. 
The cartoon also shows that the green and orange QSLs are connected to each other 
via a third QSL whose footprints have a shape very similar to the QSL of the null-point 
configuration studied in \cite{Masson09} and \cite{Reid12}.

While we did not find a null-point associated with $Q_3$, the topology of 
the magnetic field in the region of $Q_{3,circ}$, as well as the corresponding 
circular flare-ribbons, are typical signatures of the presence of a magnetic null-point 
\citep[see \eg][]{Masson09,Wang12,Deng13}. The circular-like shape of the positive 
magnetic polarity in this region and the low negative magnetic flux suggest the 
presence of a very low-lying, nearly photospheric null-point. 
The absence of a null-point in the corresponding region of our 
LFFF extrapolation is very likely related 
to the strength of the magnetic field measured by HMI. Indeed, in the region of 
$Q_3$, the HMI data display three distinctive negative magnetic polarities whose 
magnetic field is of the order of $\lessapprox 9 \ \textrm{Gauss}$, which is lower 
than the $10 \ \textrm{Gauss}$ of HMI sensitivity. We conjecture that the absence of 
a null-point in the corresponding region of our LFFF extrapolation is not inherent to 
the extrapolation, but is due to a poor precision in the measurement of the weak negative 
flux --- whose strength is comparable to the instrument sensitivity --- which prevents from 
retrieving the null.

Overall, the above results show that AR 11589 presents a complex topology which 
comprises two double {C-shaped} QSLs, one quasi-separator that links them both 
\citep[][]{Parnell10a}, and a possible null-point. Such a topology is favorable for 
the build-up of electric current layers at any of the identified QSLs 
\citep[\eg][]{Aulanier05,Haynes07}. Furthermore, any disturbance of any of these 
topological systems is likely to trigger magnetic reconnection at all the others 
\citep[\eg][]{Parnell08,Parnell10b}.

%

\section{A confined flare above filaments} \label{sec:S-confined-flare}

\subsection{Driver}  \label{sec:S-Driver}

  \begin{figure*}
\sidecaption
\includegraphics[width=0.7\textwidth,clip=]{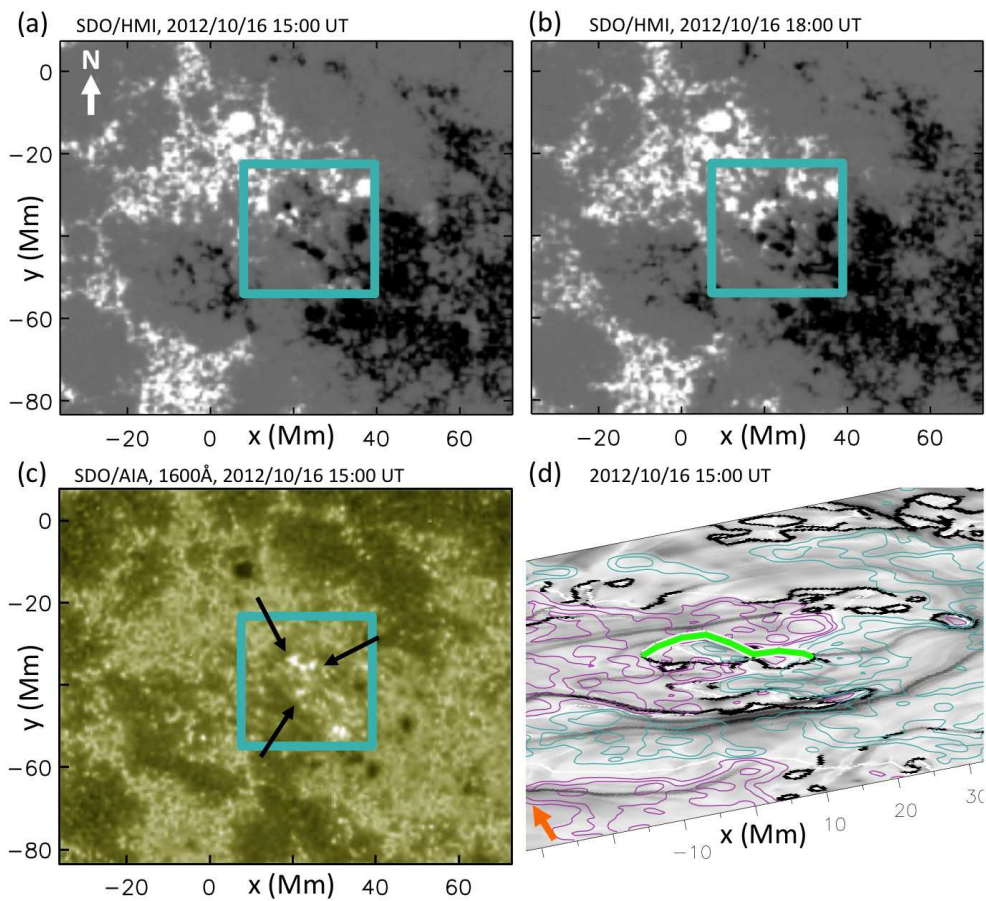}
              \caption{Signatures of magnetic flux emergence occurring in NOAA 11589 around the time of the flare. {The cyan rectangle highlights the region of magnetic flux emergence.} HMI magnetograms in greyscale (a) before the flare, and (b) after the flare. \kev{The temporal evolution of the magnetograms is available as a movie in the online edition.} (c) AIA 1600 \AA\, image showing some EBs which are highlighted by the black arrows. (d) {extrapolated serpentine field line (green) associated with the EBs shown panel (c),} plotted over the photospheric mapping of the QSLs (greyscale). Solid purple/cyan lines show {the same $B_z$ isocontours as in \fig{Fig-Bz-AIA171}.} The {white and orange arrows indicate} the north direction.
                      }
   \label{fig:Fig-Emergence}
   \end{figure*} 

To identify the possible driver of the observed C3.3 flare, we considered 
HMI and AIA 1600 \AA\, data sets {at a 12-min cadence} within a range 
of 4 hours prior to, and after, the flare.

Before the flare, the region of the magnetogram enclosed by the cyan rectangle in 
\fig{Fig-Emergence}a displayed spatially-aperiodic successions of opposite magnetic 
polarities in directions oriented from the north-east towards the south-west. 
These patterns were spatially correlated with Ellerman bombs \citep[EBs;][]{Ellerman17} 
as highlighted in \fig{Fig-Emergence}c. EBs are small recurring brightenings often 
observed in the photospheric wings of chromospheric lines 
\citep[\eg][]{Vorpahl72,Kurokawa82,Qiu00,Georgoulis02,Bernasconi02,Pariat04,Pariat07,Fang06,BelloGonzalez13,Vissers13}. 
They are believed to be the result of bald-patch reconnection occurring along 
undulatory, or serpentine, flux tubes as they cross the photosphere and emerge 
into the solar corona \citep[see][]{Pariat04,Cheung10}. 
Our LFFF extrapolation suggests that such serpentine flux tubes were indeed 
present prior to the flare, {in the region hosting EBs,} as shown \fig{Fig-Emergence}d. 
Finally, \fig{Fig-Emergence}b shows that a new bipole appeared some time after 
the flare, as inferred from the broad patches of opposite polarities present in the 
center of the cyan rectangle and which are accompanied with small-scale bipolar 
patches.

Such observational features are clear signatures of magnetic flux emergence 
starting hours before the flare onset. This emergence occurred below the QSL 
$Q_1$, in between the western part of the $Q_{1,curv}$ and the southern part 
of the $Q_{1,arc}$ branches. Furthermore, this region below $Q_1$ corresponds 
to the location of the first flare brightennings. So this continuous emergence 
below $Q_1$ may well have induced magnetic reconnection at this QSL. It may 
thus have been responsible for the trigger of the flare 
\citep[\eg][]{Schmieder97,Bagala00,DelZanna06}. 
Hence, we conjecture that continuous emergence starting prior to the flare, 
and occurring below the northern QSL of the AR, was the driver of the observed 
C-class flare.

\subsection{{Proposed} flare scenario} \label{sec:S-Flare-scenario}

We propose that the observed C-class flare was the result of a multiple-step 
reconnection mechanism driven by magnetic flux emergence below $Q_1$. 
{In this scenario, the continuous magnetic flux emergence below $Q_1$ leads 
to the accumulation of magnetic stress at $Q_1$, which results in the build-up of an 
electric current layer at this QSL \citep[\eg][]{Milano99,Aulanier05,Torok09}. This emergence 
leads to the intensification and the thinning of this current layer, which eventually 
triggers slipping/slip-running magnetic reconnection \citep{Aulanier06b}, at $Q_1$, 
of the emerging field with the ambient pre-existing magnetic field.

Because of the proximity of $Q_{1,arc}$ with $Q_{2,curv}$ and $Q_{3,curv}$, or $Q_{1,curv}$ 
with $Q_{2,arc}$ and $Q_{3,circ}$, the slipping/slip-running magnetic reconnection 
at $Q_1$ is likely to stress the magnetic field of $Q_{2}$ and $Q_{3}$ since magnetic 
stress can be transported at all QSLs via the quasi-separator that links the QSLs 
all together \citep[\eg][]{Priest96,Galsgaard97,Parnell08}. 
Indeed, at the quasi-separator, the QSLs share common magnetic field lines. 
The stress of such field lines at one of the QSLs is thus likely to also build-up 
stress at the quasi-separator and/or at the other QSL(s) sharing these field lines. 
Such a stress may build-up electric currents at $Q_{2}$ and $Q_{3}$, or may 
increase pre-existing electric currents within these two QSLs. Eventually, the induced 
stress of $Q_2$ and/or $Q_3$ triggers magnetic reconnection at these two QSLs.

In our scenario, the flare is thus the consequence of continuous slow emergence 
of magnetic flux below $Q_1$, which results in slipping/slip-running reconnection 
at this QSL, eventually triggering reconnection at the two other interlinked QSLs. 
Particle acceleration is thus expected at all QSLs, implying the formation of 
flare-ribbons at all QSLs footprints, and post-flare loops anchored into the flare-ribbons 
\citep[\eg][]{Gorbachev89,Schmieder97,Mandrini06,Baker09,Chandra11}, 
as supported by the AIA 1600 and 193 \AA\, data in 
\figs{Fig-QSLs-AIA1600}{Fig-193-pfl} in this particular event.

It must be emphasized that all three QSLs involved in our flare scenario are 
located above the two observed non-eruptive filaments which are passive during 
the flare that spreads in the corona {\it above} and {\it around} them. 
This {\it a posteriori} supports the assumption made in \sect{S-Bextrapol}, that the 
flare mechanism did not involve the magnetic field of the filaments. 
Our scenario thus explains the formation of the two extended flare-ribbons around 
the two filaments, as the consequence of sequential magnetic reconnection occurring 
in a complex system of three interlinked QSLs located above the filaments.
}

%

\section{Summary and Discussion} \label{sec:S-Summary-Discussion}

%
%
In this study, we used multi-wavelength, high-resolution observations obtained by 
the SDO, ARIES and THEMIS instruments, so to analyze the dynamics of the magnetic 
field of AR NOAA 11589 that led to a non-standard C3.3 class flare on 2012 October 
16. The AR evolution was associated with large-scale magnetic flux cancellation that 
led to the formation of two filaments of opposite chirality. Unlike what the standard 
model predicts, the flare loops formed {\it above} and not {\it below} the filaments. 
Furthermore, the latter were apparently not involved in the flare mechanism, since 
they did not erupt. The dataset considered here also presented the signatures of 
localized magnetic flux emergence episodes in the northern part of the AR. Our 
analysis indicates that the flare was driven by one of these episodes that actually 
took place below a complex system of quasi-separatrix layers (QSLs), as calculated 
in a linear force-free field (LFFF) extrapolation. This continuous magnetic flux emergence 
presumably stressed the magnetic field of the QSLs, thus resulting in the development 
of narrow and intense current layers within them. This scenario implies the occurrence 
of multiple and sequential magnetic reconnections within the complex set of QSLs, 
which led to the observed flare. This scenario is supported by the relatively good 
match found between the expected timing of the QSL activations, the shape of the 
QSL footprints, and the development and morphology of complex flare ribbons and 
loops as observed in the EUV {(see online movies associated with 
\figs{Fig-flare}{Fig-193-pfl}).}

%
%
{By performing a set of LFFF extrapolations using different values of the force-free 
parameter, we have demonstrated the robustness of the derived complex topology, 
and hence of our results. 
More generally, our study shows the stability of the QSLs related to large-scale coronal 
loops/magnetic fields that are not associated with a magnetic flux-rope. In particular, 
it shows the stability of such QSLs (1) against 
changes --- within a certain range --- of the force-free parameter for LFFF extrapolations 
\citep[see also][]{Aulanier05}, and (2) against temporal variations 
that do not result in a major evolution of the photospheric magnetic flux and/or of electric 
currents \citep[see also the large-scale QSL of the quadrupolar AR 11158 in][]{Zhao14}. 
We recall that the force-free parameter controls the amount of electric current density 
in magnetic field lines, which can be observationally related to the 
photospheric transverse/horizontal magnetic field. Therefore, the stability of the QSLs of 
large-scale coronal loops/magnetic fields --- that are not associated with 
a magnetic flux-rope --- suggests that such QSLs are mainly constrained by 
the photospheric longitudinal/vertical 
magnetic field, hence, by the large-scale distribution of the photospheric magnetic flux.}

%
%
It is worth noticing that the flare scenario that we proposed is based on one important 
conjecture, namely that slip-running reconnection may activate several QSLs which 
are linked together. This may be expected because reconnecting field lines may slip 
from one QSL to another. In this picture, a given field line may reconnect at least two 
times in the considered magnetic configuration. Such sequences of magnetic 
reconnections for a given field line have already been reported for magnetic 
configurations with separatrices intersecting at a separator 
\citep[\eg][]{Galsgaard97,Haynes07,Parnell10a}. However, to the authors' knowledge, 
it has never been shown to occur in complex QSL systems in which two {QSLs} 
are located in the vicinity of one another. Therefore, this conjecture 
should be addressed by future numerical experiments in which the initial magnetic 
field configurations should possess two neighboring {QSLs.}

%
%
The C3.3 class flare analyzed in this paper is a typical example of an atypical 
flare exhibiting signatures common to both standard and confined solar flares. 
Indeed, at large scales, the flare initially appears to be associated with the formation 
of two {extended} ribbons that developed parallel to and aside the filaments, 
in a globally bipolar 
active region, just like in the standard model. However, at smaller scales, the polarity 
inversion line is strongly curved. The ribbons have a complex shape, and they did not 
brighten simultaneously. Together, these two features suggest some coupling of remote 
regions that did not seem to be magnetically linked to the filaments. Furthermore, the 
filaments did neither erupt, nor were they associated with any failed eruption. 
Explaining this type of atypical events in general may be a challenge for the usual 
eruptive and confined flare models. Nevertheless, the topological analysis of the 
magnetic field derived from a force-free extrapolation, here achieved using the QSL 
method \citep[{applied with the squashing degree, Q} ;][]{Demoulin97}, shows 
that it is possible to explain atypical flare signatures 
as a complex QSL system which allows to couple remote regions via slip-running 
reconnection \citep{Aulanier06b}.

On the one hand, this work further confirms that QSLs 
play a key role for 3D reconnection in solar flares, as reported in previous studies of 
less complex events \citep[\eg][]{Schmieder97,Mandrini06,Chandra11}. On the other 
hand, this study suggests that topological analyses, such as the QSL method 
{(using either N or Q),} may also be the answer to explaining atypical solar flares, 
that may actually be more numerous than the more classical eruptive and confined 
flares which are often analyzed in the literature. This conclusion is further 
confirmed by the topological analysis of a different atypical flare studied, in the 
framework of the QSL method, by \cite{Liu14}. 
In their event, the magnetic configuration was derived using a nonlinear force-free 
field (NLFFF) extrapolation. Similarly to our event, the derived configuration possessed 
a large-scale QSL above a magnetic flux-rope (although our event was associated 
with two QSLs and two filaments, each QSL lying above a filament). As in our case, 
the flare was likely driven by magnetic flux emergence occurring below the large-scale 
QSL, in a region different from the flux-rope location, and which eventually triggered 
magnetic reconnection at this QSL. However, contrary to our event, the continuous 
reconnection at the large-scale QSL of their configuration eventually destabilized 
the flux-rope whose eruption failed due to the presence of strong confining arcades 
above it.

%
%
If atypical solar flares are the most numerous, then the study by \cite{Liu14} and ours 
suggest that the classical paradigm of confined and eruptive flares should be 
revisited. Note, however, that these are \kev{only two independent case studies,} 
so further topological analyses of atypical solar flares, either using LFFF or NLFFF 
extrapolations, are required to confirm such a statement and that topological studies 
are indeed relevant for all these complex events.

\begin{acknowledgements}
We thank the referee for helpful comments that improved the paper. 
We deeply thank Dr. W. Uddin for providing us with the observations of the ARIES 
telescope. We thank all the team of THEMIS for adjusting the telescope during our 
observing campaign and the director, Bernard Gelly for providing us with the data. 
K.D. thanks E. Pariat for fruitful discussions that helped in identifying the 
driver of the flare. R.C. thanks the Observatoire de Paris for the grant given during 
his stay in Meudon in January 2013. We acknowledge the open data policy of 
NASA/SDO.
\end{acknowledgements}

\bibliographystyle{aa}
      
\bibliography{Flare_NOAA11589_Dalmasse_etal}  

\IfFileExists{\jobname.bbl}{} {\typeout{}
\typeout{****************************************************}
\typeout{****************************************************}
\typeout{** Please run "bibtex \jobname" to obtain} \typeout{**
the bibliography and then re-run LaTeX} \typeout{** twice to fix
the references !}
\typeout{****************************************************}
\typeout{****************************************************}
\typeout{}}

\end{document}